\documentclass[a4paper,12pt]{article}
\usepackage{epsfig}
\date{\today}
\newcommand{\bmat}{\left(\begin{array}}
\newcommand{\emat}{\end{array}\right)}

\def\npb#1#2#3{    {\it Nucl. Phys. }{\bf B#1} (#2) #3}

\def\plb#1#2#3{    {\it Phys. Lett. }{\bf B#1} (#2) #3}
\def\prd#1#2#3{    {\it Phys. Rev. }{\bf D#1} (#2) #3}

\def\prl#1#2#3{    {\it Phys. Rev. Lett. }{\bf #1} (#2) #3}

\def\Ibanez{Ib\'a\~{n}ez~}
\def\Munoz{Mu\~{n}oz~}

\def\lsim{\raise0.3ex\hbox{$\;<$\kern-0.75em\raise-1.1ex\hbox{$\sim\;$}}}
\def\gsim{\raise0.3ex\hbox{$\;>$\kern-0.75em\raise-1.1ex\hbox{$\sim\;$}}}

\def    \part          {\partial}
\def    \be            {\begin{equation}}
\def    \ee            {\end{equation}}
\def    \bea           {\begin{eqnarray}}
\def    \eea           {\end{eqnarray}}
\def    \nn            {\nonumber}

\begin{document}

\renewcommand{\thefootnote}{\fnsymbol{footnote}}

\begin{titlepage}
\pagestyle{empty}
\rightline{SUSX-TH-024}
\rightline{FISIST/21-2000/CFIF}
\rightline{January 2001}
\vskip 1cm
\begin{center}
{\bf \large{Lepton--Flavor Violation with Non--universal Soft
Terms\\[10mm]}}
{D.F. Carvalho$^1$\footnote{dani@gtae2.ist.utl.pt}, 
M.E. G\'omez$^1$\footnote{mgomez@cfif.ist.utl.pt}, 
and  S. Khalil$^{2,3}$\footnote{s.khalil@sussex.ac.uk}\\[6mm]}
\small{
$^1$ CFIF, Departamento de F\'{\i}sica,
Instituto Superior T\'ecnico, Av. Rovisco
Pais,\\[-0.3em] 1049-001 Lisboa, Portugal\\
$^2$ Centre for Theoretical Physics, University of Sussex, Brighton BN1
9QJ,~~U.~K.\\
$^3$ Ain Shams University, Faculty of Science, Cairo, 11566, Egypt.\\[7mm]}
\small{\bf Abstract}
\\[7mm]
\end{center}
\begin{center}
\begin{minipage}[h]{14.0cm}
We study the lepton--flavor violation processes $\tau \to \mu \gamma$ and
$\mu \to e \gamma$ in two different examples of models with non--universal
soft breaking terms derived from strings. We show that the
predictions are quite different from those of universal scenarios.
Non--universal $A$--terms provide an interesting framework to
enhance the supersymmetric contributions to CP violation
effects. We observe that in the case of the lepton--flavor violation
we study, the non--universality of the scalar masses 
enhances the branching ratios more significantly than the
non--universality of the $A$--terms. We find that the
current experimental bounds on these processes restrict both the
parameter space of the models and the texture of the Yukawa
couplings which predicts the lepton masses, providing at the same time an
interesting experimental test for physics beyond the Standard
Model.

\end{minipage}
\end{center}
\end{titlepage}
\newpage
\setcounter{page}{1}
\pagestyle{plain}
\renewcommand{\thefootnote}{\arabic{footnote}}
\setcounter{footnote}{0}
%
\section{{\bf \large Introduction}}

Flavor changing neutral currents (FCNC's) provide an
important test for any new physics beyond the Standard Model (SM).
It is well known that in the SM both baryon and lepton numbers 
are automatically
conserved and  that tree--level flavor--changing neutral currents are absent.
After the recent result from Super--Kamiokande, which is regarded as 
compelling evidence for the oscillation of $\nu_{\mu}$ to $\nu_{\tau}$ with a
squared mass difference of order $10^{-3}-10^{-2}$
eV$^2$~\cite{komiokande,chooz}, there has been crescent interest
in lepton--flavor violation (LFV). The rates for charged LFV processes 
are extremely
small in the SM with right--handed neutrinos ($\propto \Delta
m_{\nu}^2/M_W^2$~\cite{P..}). The present experimental
limits~\cite{lfv-limit} are
\bea
BR(\mu \to e \gamma) &<& 1.2 \times 10^{-11} \nn\\
BR(\tau \to \mu \gamma) &<& 1.1 \times 10^{-6} \nn\\
BR(\tau \to e \gamma) &<& 2.7 \times 10^{-6}.
\label{limit}
\eea
Thus, the observation of these processes would be a
signal of new physics. In different extensions of the SM, predictions of
LFV processes compatible with the above experimental limits have been studied.
For instance, supersymmetric (SUSY) models with a gauge unification group
(GUT) and  SUSY models with ``see-saw'' neutrinos
have been discussed in Refs.~\cite{rh-nu,hisano,mg,u1-sy},
SUSY models with $R$--parity violation in Ref.~\cite{r-par}, and recently, 
models with extra dimensions in Ref.~\cite{extra}. LFV can be observed
in other systems, such as $\mu\to e$ conversion on heavy nuclei
and $\mu\to 3 e$ \cite{hisano,mg}, rare kaon decays~\cite{kaon}, and
slepton flavor mixing at future accelerator experiments
\cite{acc}. However, the first two processes listed above
are the most restrictive in the models considered in this work; the
predictions for $ BR(\tau \to e \gamma)$ are typically much lower
than the bound~(\ref{limit}).

In R--parity conserving SUSY models, the presence of LFV processes is
associated with vertices involving leptons and their
superpartners~\cite{oldflav}. In the Minimal Supersymmetric Standard Model
(MSSM), with  universal soft terms,  
it is possible to rotate the charged lepton Yukawa couplings
and the sleptons in such a way that lepton flavor is preserved. However,
a small deviation from flavor universality in the soft-terms at the GUT
scale will be severely constrained by the experimental
bounds~(\ref{limit}). In fact GUT theories ~\cite{rh-nu}
and models with $U(1)$~family symmetries~\cite{u1-sy}
can lead to the  MSSM with flavor--dependent soft terms leading to
important violations of the lepton flavor.

String--inspired models naturally lead to such non--universality
in the soft SUSY breaking sector~\cite{ibanez1, ibanez2, gordy1},
and in such models a non--diagonal texture for the Yukawa couplings 
will generate a 
flavor mixing soft--term structure.
Therefore, the LFV predictions will impose additional constraints
on the free parameters of this kind of models. In this work we
concentrate ourselves on  two examples of string--inspired models with
non--universal soft terms. The relevance of the non--universality
of the trilinear terms ($A$--terms) has recently been considered
in Refs.~\cite{khalil,tatsuo, gordy2}.
It was shown that the flavor structure of the $A$--terms is
crucial for enhancing the SUSY contributions to CP violation
effects, and for generating the experimentally observed $\varepsilon$
and $\varepsilon'/\varepsilon$. These models also 
predict a CP asymmetry in $B \to X_s \gamma$ decay much
larger than the one predicted by SM and the one obtained for 
a wide
region of the parameter space in 
minimal supergravity scenarios~\cite{bailin}.
It has also been noted
that the analysis of $b \to s \gamma$ does not severely constrain
the models under consideration~\cite{emidio}. However, 
the non--universality of the
$A$--terms in this class of models is always associated with  
non--universal  scalar masses. In this case a simple 
non--diagonal Yukawa texture predicting lepton masses will induce 
two sources of
LFV: one due to the flavor structure of the $A_l$--terms which
prohibits the simultaneous diagonalization of the lepton Yukawa matrix
$Y_l$ and the trilinear couplings $(Y_l^A)_{ij}\equiv (A_l)_{ij}
(Y_l)_{ij}$; the other source is due to the non degeneracy of
the scalar masses of the sleptons. Therefore, in the basis where $m_l$ is
diagonal the slepton mass matrix acquires non--diagonal
contributions. We find that in general the second source dominates
over the first in the case of the LFV predictions.

We analyze the dependence of our results on the lepton Yukawa texture.
We study two different $Ans\ddot{a}tze$ for $Y_l$, leading to the 
correct prediction
for the charged lepton masses. The form of  $Y_l$ can be further specified 
when a predictive mechanism for the
neutrino sector is included in the models. However we do not address 
this issue here. As we will see, the structure of $Y_l$ is decisive in
determining to what extent the soft terms generated by the models can deviate 
from the universal case.

The paper is organized as follows. In section 2 we present the
structure of the soft terms in the two string models we analyze. In 
section 3 we present the two examples of lepton Yukawa couplings and the
slepton mass matrices.
In section 4 we present the theoretical
framework for calculating the LFV in this class of models.
Our predictions for charged leptonic rare processes are given in section 5 for
weakly coupled heterotic string (WCHS), and those for the type I string
model are given in section 6. Finally we give
our conclusions in section 7.

\section{{\bf \large Flavor structure of soft SUSY breaking parameters}}
In this section we briefly discuss the possible mechanisms
which may give rise to the
non--universal soft SUSY breaking terms. 
As already mentioned, 
motivated by the minimal supergravity model, it is
common to assume, that the soft
SUSY breaking terms are universal
at the GUT scale. 
However, it is possible to obtain  effective
potentials in which this
universality is absent, as it is the case of considering the kinetic terms 
for the chiral superfields to be non--minimal. As recently
stressed in Refs.~\cite{ibanez1,ibanez2}, the soft SUSY breaking
parameters may be non--universal in the effective theories derived from 
superstring theories.

The problem of SUSY breaking is not completely understood yet, even in 
superstring theories. However, generic superstring models include a dilaton 
field $S$ and moduli fields $T_i$, these are gauge singlet fields with  
their coupling to the gauge non--singlet matter being 
suppressed by powers of Planck mass. Therefore, they can naturally 
constitute a `hidden sector'. Recently~\cite{ibanez1, ibanez2}, the soft SUSY 
breaking terms have been derived under the assumption 
that SUSY is broken only by the vacuum expectation values (VEV's) of 
$F$--terms corresponding to these $S$ and $T_i$ fields.

This general framework has been used to study the implications
on the effective supergravity theories  which emerge in the low 
energy limit of the weakly coupled 
heterotic strings (WCHS)~\cite{ibanez1} and type I
string models~\cite{ibanez2}. These two examples show that the soft
SUSY breaking terms are, in general, non--universal. The details 
for deriving these soft terms are given in Ref.~\cite{ibanez1,ibanez2} and 
some aspect of their phenomenological implications can be 
found in Ref.~\cite{khalil}. Here we briefly present the soft 
terms which are essential for our work.

In the WCHS case,  it is assumed that the superpotential of
the dilaton ($S$) and moduli ($T$) fields is
generated by some non--perturbative mechanism, and that
the $F$-terms of $S$ and $T$ contribute to the SUSY breaking.
Hence one can parametrize the $F$-terms as~\cite{ibanez1}
\be
F^S = \sqrt{3} m_{3/2} (S+S^*) \sin\theta,\hspace{0.75cm} F^T
=m_{3/2} (T+T^*) \cos\theta .
\ee
Here $m_{3/2}$ is the gravitino mass and $\tan \theta$ corresponds to the ratio between the $F$-terms of $S$ and $T$.
In this framework, the soft scalar masses $m_i$ and the gaugino masses
$M_a$ are given by~\cite{ibanez1}
\begin{eqnarray}
m^2_i &=& m^2_{3/2}(1 + n_i \cos^2\theta), \label{scalar}\\ M_a
&=& \sqrt{3} m_{3/2} \sin\theta ,\label{gaugino}
\end{eqnarray}
where $n_i$ is the modular weight of the corresponding field. The $A$-terms are written as
\begin{eqnarray}
A_{ijk} &=& - \sqrt{3} m_{3/2} \sin\theta- m_{3/2}
\cos\theta (3 + n_i + n_j + n_{k}), \label{trilinear}
\end{eqnarray}
where $n_{i,j,k}$ are the modular weights of the fields
that are coupled by this $A$--term.
If we assume $n_i=-1$ for the third family and $n_i=-2$
for the first and second
families (in addition we take  $n_{H_1}=-1$ and $n_{H_2}=-2$), we find 
the following
texture for the $A_l$-parameter matrix at the string scale
\begin{equation}
A_l = \left (
\begin{array}{ccc}
x & x & y\\
x & x & y \\
y & y & z
\end{array}
\right),
\label{AtermA}
\end{equation}
where
\begin{eqnarray}
x&=& m_{3/2}(-\sqrt{3} \sin\theta + 2  \cos\theta),\\
y&=& m_{3/2}(-\sqrt{3} \sin\theta + \cos\theta),\\
z&=&-\sqrt{3}m_{3/2}\sin\theta.
\end{eqnarray}

Our choice of modular weights is motivated from the fact that assigning 
different values of the modular weights for the first and  
second families would make their scalar masses non degenerate,
this would imply in general  values for $BR(\mu\to e \gamma)$ 
in conflict with the  experimental limit (\ref{limit}). Conversely, 
assigning a common modular weight for
all the families would lead to degenerate scalar masses as well as 
universal $A$--terms and hence lepton flavor would be preserved.

The deviation from the universality of the soft--terms in this  model can be 
parameterized by the angle $\theta$. The value $\theta =\pi/2$
corresponds to the universal limit for the soft terms. In order to
avoid negative  squared  values in the scalar masses we restrict
ourselves to the case with $\cos^2 \theta < 1/2$. Such
restriction on $\theta$ makes the deviation from the 
universality in the whole
soft SUSY breaking terms very limited. However, as shown in
\cite{khalil}, this small deviation from the universality of the soft terms 
is enough to generate sizeable SUSY CP violations
in the  $K^0-\overline{K}^0$ system.

In type I string model case, non--universality
in the scalar masses, $A$--terms and gaugino masses
may be naturally obtained~\cite{ibanez2}.  Type I models can contain
9--branes, $5_i$--branes, $7_i$--branes and 3 branes where the index $i = 1,2,3$
denote the complex compact coordinate which is included in the $5$--brane world 
volum or which is orthogonal to the $7$--brane world volume. However, to 
preserve $N=1$ supersymmetry in $D=4$ not all of these branes can be present 
simultaneously and we can have either $D9$--branes with $D5_i$--branes or
$D3$--branes with $D7_i$--branes. From the phenomenological point of view there
is no difference between these two scenarios. Here we consider the
model used in Ref.~\cite{tatsuo}, where the gauge group
$SU(3)_C \times U(1)_Y$ is associated with the 9--brane while $SU(2)_L$
is associated with the $5_1$--brane.

The SUSY breaking is analyzed, as in WCHS model,
in terms of the VEV's of the dilaton and moduli fields \cite{ibanez2}
\be
F^S = \sqrt{3} m_{3/2} (S+S^*) \sin\theta,\hspace{0.75cm} F^{T_i}
=m_{3/2} (T_i+T_i^*) \Theta_i \cos\theta~,
\ee
where $i=1,2,3$, 
the angle $\theta$ and the parameters $\Theta_i$ just parametrize 
the direction of the
goldstino in the $S$ and $T_i$ fields space. The parameters $\Theta_i$ 
verify the relation,
\be
\sum_i \left|\Theta_i\right|^2=1,
\label{sumteta}
\ee
Within this framework, the gaugino masses are~\cite{ibanez2}:
\bea
M_1 &=& M_3 = \sqrt{3} m_{3/2} \sin\theta ,\label{m1}\\
M_2 &=& \sqrt{3}m_{3/2} \Theta_1 \cos \theta .\label{m2}
\eea
In this model the fermion doublets and the Higgs fields are assigned to
the open string which spans between the $5_1$--brane and $9$--brane,
while the fermion singlets correspond to open strings which start
and end on the $9$ brane. Such open string states include three sectors
which correspond to the three complex compact dimensions. If we
assign the fermion singlets to different sectors we obtain
non--universal $A$--terms. It turns out that in this model the
trilinear couplings are given~\cite{ibanez2,tatsuo} by:
\begin{equation}
A_u=A_d=A_l= \left (
\begin{array}{ccc}
x & y & z\\
x & y & z \\
x & y & z
\end{array}
\right),
\label{AtermB1}
\end{equation}
where
\begin{eqnarray}
x &=& - \sqrt{3} m_{3/2}\left(\sin\theta + (\Theta_1 - \Theta_3) \cos\theta
\right),\\
y &=& - \sqrt{3} m_{3/2}\left(\sin\theta + (\Theta_1 - \Theta_2) \cos\theta
\right),\\
z &=& - \sqrt{3} m_{3/2} \sin\theta.
\label{AtermB2}
\end{eqnarray}
The soft scalar masses for sfermion-doublets $(m^2_L)$, and the 
sfermion-singlets $(m^2_{R_i})$ are given by
\bea
m^2_L &=&  m_{3/2}^2 \left( 1- \frac{3}{2} (1-\Theta_1^2) \cos^2
\theta\right)\label{scalarBL} ,
\label{sSL}
\\ m^2_{R_i} &=&  m_{3/2}^2 \left( 1- 3 \Theta_i^2 \cos^2
\theta\right),
\label{scalarBR}
\eea
where $i = 1,2,3$ refers to the third, second and first family respectively 
refers to the three families \cite{tatsuo}. The soft masses for the 
Higgs fields are similar to the one of the   sfermion-doublets 
(\ref{sSL}).
For $\Theta_{i} = 1/\sqrt{3}$ the $A$--terms and the
scalar masses are universal while the gaugino masses can  be
non--universal. The universal gaugino masses are obtained for 
$\theta=\pi/6$.

\section{\bf \large Slepton mass matrices and Yukawa textures}

In order to completely specify the models described in the previous section,
we have to fix the Yukawa textures and hence determine the flavor
structure of the slepton mass matrices.

In Refs.~\cite{khalil}, some phenomenological consequences of the 
flavor dependence of the squark soft terms were studied.
These works assume some typical 
quark Yukawa textures with satisfactory predictions 
for quark masses and mixings. Similarly,  a general Yukawa texture for the 
leptonic sector will translate the flavor dependence of the soft
terms at the GUT scale into flavor mixing lepton--slepton vertices.

We should emphasize that the only experimental constraint on the lepton 
Yukawa couplings in the context of the MSSM with flavor blind soft terms 
(i.e. as derived from minimal supergravity) is the correct prediction
for the lepton masses. However, with soft terms as described 
in the previous section, the results stated in the present work
will depend strongly on the structure of the Yukawa texture assumed for
the lepton sector.

To illustrate the dependence of the next sections results on the 
lepton Yukawa couplings, we consider 
two examples of symmetric textures at the GUT scale:

\begin{itemize}
\item Texture I,
\be
Y_{l}=y^{\tau }\left( \begin{array}{ccc}
0 & 5.07\times 10^{-3} & 0\\
5.07\times 10^{-3} & 8.37\times 10^{-2} & 0.4\\
0 & 0.4 & 1
\end{array}\right)
\ee 
\item Texture II,
\be
Y_{l}=y^{\tau }\left( \begin{array}{ccc}
3.3\times 10^{-4} & 1.64\times 10^{-5} & 0\\
1.64\times 10^{-5} & 8.55\times 10^{-2} & 0.4\\
0 & 0.4 & 1
\end{array}\right)
\ee
\end{itemize}

Both of them lead to the correct prediction for the experimental values of 
the lepton masses.

Texture I is a symmetric texture used in some of the  solutions
obtained in ~\cite{RRR}. Our motivation is
that this texture can be considered to be the limiting case of
textures arising from  $U(1)$~family symmetries as described in 
Refs.~\cite{u1m}
and studied in Refs.~\cite{mg} in the context of LFV induced 
by R-H neutrinos.
As we will see, when this texture is considered,
the BR( $\mu\to e \gamma$) imposes a severe constraint on the parameter 
space of the models, allowing a very small deviation from the
universality on the soft terms. Typically a prediction for the decay
 $\tau \to \mu \gamma$ of the order of the experimental limit~(\ref{limit})
will imply a severe violation of the experimental bound for
 $\mu \to e \gamma$.

Texture II provides a good prediction for the lepton masses and induces 
branching ratios  $\mu \to e \gamma$ and $\tau \to \mu \gamma$ of the
order of the current  experimental bounds. This texture was chosen as an 
illustration of how the current bounds (\ref{limit}) can provide 
some information about the lepton Yukawa couplings on the context of the 
models considered.

We should stress that texture I can fit in 
a more complete model of Yukawa matrices, as the ones aimed to 
explain fermion  masses and 
quark mixings with a minimal amount of input parameters.
On the other hand, we selected texture II based on the predictions for the
processes under consideration. It is beyond the purpose of our work to 
include this texture in the context of a general model for the
Yukawa couplings. However, GUT theories such as $SU(3)_c\times SU(3)_L
\times SU(3)_R$ can  provide an example of lepton Yukawa couplings not
related to those of the quarks (see for example Ref.~\cite{LR}). We 
must also observe that unlike
texture I, the structure of texture II cannot be similar to the ones obtained
in the models with $U(1)$--family symmeties as described 
in Refs.~\cite{u1m,LR}, since it is not possible to 
arrange the $U(1)$--charges in order to have the element 
$(1,1)$ of the texture larger than the $(1,2)$ and $(1,3)$ elements.

The texture assumed for the
quark Yukawa couplings has a marginal effect on the computation of
the branching ratios studied in the present work. Nevertheless, we 
need the 
full matricial structure of  all Yukawa couplings in order to be 
consistent with our analysis of the renormalization--group 
equations (RGE's). Therefore, we use the  
$Ans\ddot{a}tze$ given in  solution 2 of Ref.~\cite{RRR}, 
with inputs at the GUT scale leading to the correct experimental
predictions for the quark sector.

The lepton Yukawa couplings can be diagonalized by the
unitary matrices $U_L$ and $U_R$ as follows,
\be
m_l = \frac{v \cos
\beta}{\sqrt{2}} U_R~ (Y^l)^T~ U^{\dag}_L .
\ee
When the superfields are written in this basis, the 
expressions for the charged
slepton mass matrices at low energy take the form:
\bea
\begin{array}{lr}
M^2_{\tilde{l}}=
\left(\begin{array}{cc}
\left(M^2_{\tilde{l}}\right)_{LL}
&\left(M^2_{\tilde{l}}\right)_{LR}
\\
\left(M^2_{\tilde{l}}\right)_{RL}
&\left(M^2_{\tilde{l}}\right)_{RR}
\end{array} \right),
\end{array}
\eea
where,
\bea
\left(M^2_{\tilde{l}}\right)_{LL}&=&U_L m^2_{L} U_L^{\dag}
+m_l^2 -\frac{m_Z^2}{2}(1-2\sin^2{\theta_W})\cos{2\beta},\nn \\
\left(M^2_{\tilde{l}}\right)_{RR}&=&U_R (m^2_{R})^T U_R^{\dag}
+m_l^2  + m_Z^2 \sin^2{\theta_W}\cos{2\beta},\nn \\
\left(M^2_{\tilde{l}}\right)_{LR}&=&\left(M^2_{\tilde{l}}\right)_{RL}^{\dag}=
-\mu~m_l \tan{\beta}+\frac{v\cos{\beta}}{\sqrt{2}}
U_L Y^{A *}_l U_R^{\dag} ,
\eea
where $m^2_{L}$ and $m^2_{R}$ are the soft breaking $(3\times3)$ mass
matrices for the charged slepton doublet and singlet respectively.

The sneutrino mass matrix is simply  given  by the $(3\times3)$ mass matrix:
\be
M^2_{\tilde{\nu}}=U_L m^2_{L} U_L^{\dag}+
\frac{m_Z^2}{2}\cos{2\beta}
\label{sneutrino}
\ee
The relevant lepton--flavor changing mass matrix elements
on the slepton  mass matrices above are given by:
\bea
(\delta^l_{LL})_{ij} &=& \left[~ U_L~ m^2_{L}
~U_L^{\dag}~ \right]_{ij}\nn \\
(\delta^l_{LR})_{ij} &=&  \left[~ U_L~ Y^{A *}_l
~U_R^{\dag}~ \right]_{ij} \label{Deltas} \\
(\delta^l_{RR})_{ij} &=& \left[~ U_R~
(m^2_{R})^T~ U_R^{\dag}~ \right]_{ij} \nn
\eea
where $i$, $j$ are flavor indices ( $i\neq j$).
We found that, in general, $\delta^l_{LL}$ and $\delta^l_{RR}$ are much
more enhanced by the non degeneracy  of the of the scalar
soft masses that what  $\delta^l_{LR}$ is due to the 
non--universality of the $A$--terms.


\section{{\large \bf LFV in SUSY models}}
Fig.~1 shows the one--loop diagrams that are relevant to the $\mu\to e \gamma$
process. The corresponding  $\tau \to \mu  \gamma$ can be represented by
analogous graphs. The amplitude for the decay can be written as a
magnetic transition:
\be
{\cal T}(l_j \to l_i \gamma)= e\ \epsilon^\lambda {\bar u_i}(p-q)
      \{ m_j i\sigma_{\lambda\beta}q^\beta
               \left(A^L P_L+A^R P_R\right)
      \} u_j(p),
\label{general}
\ee
where $q$ is the photon momentum. $A^L$ and
$A^R$ receive contributions from both neutralino--charged
slepton ($n$) and chargino-sneutrino ($c$) exchange
\begin{equation}
A^{L,R}=A_{n}^{L,R}+A_{c}^{L,R}.
\label{ampl}
\ee

In the limit of vanishing mass for the outgoing leptons,
Eq.~\ref{general} can be written as
\be
{\cal T}(l_j \to l_i \gamma)= e\ {\bar u_i}(p-q)
      \{ 2 p\cdot \epsilon \ m_j
               \left(A^L P_L+A^R P_R\right)
      \} u_j(p).
\label{general1}
\ee

Thus the decay rate is given by:
\be
\Gamma(l_j \to l_i \gamma)=\frac{m_{l_j}^3}{16 \pi}(|A_R|^2+|A_L|^2).
\label{rate}
\ee

The formulae used to calculate these amplitudes can be found in
Ref.~\cite{hisano}.
The evaluation of the branching ratios for
the $\mu\to e\gamma$ and $\tau\to \mu \gamma$ decays involve the masses
of a good part of the supersymmetric particles. Therefore, it is important 
to know precisely all masses and
other low energy parameters for any given set of inputs at the GUT scale
($M_{GUT}$). In the present work,
this is obtained by numerical integration
of the RGE's of the MSSM. The complete 
RGE's at two loops can be found in Ref.~\cite{rge}.

In addition to the soft--breaking terms dictated by the string--models under
consideration, our effective theory below $M_{GUT}$ depends 
on the parameters:
\[
 \alpha_G,\ M_{GUT},\ \tan\beta,\ Y_u,\ Y_d,\ Y_l.
\]
 The quantities $\alpha_G=g_G^2/4 \pi$ ($g_G$ being the GUT gauge
coupling constant) and $M_{GUT}$ are evaluated consistently with
the experimental values of $\alpha_{em}$, $\alpha_s$, and
$\sin^2\theta_W$ at $m_Z$. We integrate numerically the
RGE's for the MSSM at two loops
in the gauge and Yukawa couplings and at one loop in the soft terms,  
from  $M_{GUT}$ down to a
common supersymmetric threshold $M_S\approx 200 \ \rm {GeV}$. From
this energy to $m_Z$, the RGEs of the SM are used. The value of
the B and $\mu$ parameter (up to its sign) can be expressed in
terms of the other input parameters by means of the electroweak
symmetry breaking conditions. We fix the elements of the Yukawa
matrices at the GUT scale, consistently with the experimental
values for the fermion masses and the absolute values of CKM
matrix elements. Despite the fact that we use the full matrix form for all
parameters, our results do not differ significantly from the
common approach of considering diagonal Yukawa matrices and neglecting
the two lightest generations. We must observe, however, that in
the models here discussed, the trilinear terms are not
diagonal in the basis where the Yukawa couplings are, and
therefore one can not avoid  a matricial treatment for them.
Finally we keep a fixed value for $\tan\beta=10$ and set the sign of
the $\mu$ parameter to  be positive. As we can see in
previous studies \cite{hisano,mg} , the branching ratios
under consideration increase with $\tan\beta$.

\section{{\large \bf $BR(l_j\to l_i \gamma)$ in the WCHS model}}

As stated in section 2, the non--universality of the soft terms
in the WCHS arises from the different modular weight assigned to 
each family and the parameter $\theta$. Hence the non--universalities in
eqs.~(\ref{scalar}) and (\ref{AtermA}) will translate into non--trivial values
for the flavor--changing matrix elements given in eq.~(\ref{Deltas}).
Therefore due to  presence of flavor--mixing  elements in the charged
and neutral slepton mass matrices, the two
diagrams of Fig.~1 contribute to the four partial
amplitudes of eq.~(\ref{ampl}).

The choice of input values for the SUSY parameters described in section
4 is completed once we give the model--dependent values of
the soft masses and trilinear terms. In order to do that we need to fix
the values of $m_{3/2}$ and the angle $\theta$ in
eqs.~(\ref{scalar})--(\ref{AtermA}). The splitting of the
soft masses increases from $\sin\theta=1$ (which corresponds to the
universal case) to the limiting case for  $\sin \theta=1/\sqrt{2}$
(below which some  square masses become negative). Therefore we
consider as representative for the WCHS model to present  the 
variation of the  branching
ratios with  $\sin\theta$ for fixed values of  $m_{3/2}$ as 
shown in Fig. 2. For the value
$m_{3/2}=200\ \rm{GeV}$, the mass of lightest neutralino varies 
from 100 to 147 GeV, that of 
the chargino from 190 to 277 GeV, while the lightest of the staus has masses 
of  107 to 233 GeV as $\sin \theta$ ranges from $1/\sqrt{2}$ to 1. Similarly 
for $m_{3/2}=400\ \rm{GeV}$ we found $m_{\tilde{\chi}^0}=210-295 \ \rm{GeV}$,
$m_{\tilde{\chi}^+}=400-570 \ \rm{GeV}$, 
$m_{\tilde{\tau}_2}=212-470 \ \rm{GeV}$ for the same range of $\sin \theta$.

Fig.~2 shows the results of the branching ratios under consideration.
We can see how texture I (graphic on the left) tolerates small
deviations from universality of the soft terms. The experimental bound on
$BR(\mu\to e \gamma)$ is satisfied only for $\sin \theta> .96$
($m_{3/2}=200\ \rm{GeV}$) and for $\sin \theta> .91$ 
($m_{3/2}=400\ \rm{GeV}$) while for the same range on  $\sin \theta$
the corresponding prediction for $BR(\tau\to \mu \gamma)$ is well below
the experimental bound. The values of the branching ratios decrease as
we increase $m_{3/2}$ since this translates into  an
increase of the masses of the supersymmetric particles. In order
to simplify  the presentation of
our results we fix $\tan\beta=10$. However, enlarging the 
value of $\tan\beta$
increases the prediction for the branching ratios as shown for 
example in Refs.~\cite{hisano,mg}.

The results obtained using texture II (Fig.~2, graphic on the right)
allow us to start the graph at the lowest value of $\sin
\theta=1/\sqrt{2}$. As it can  be seen, the experimental bounds are
more restrictive for the $\tau \to \mu \gamma$ than for $\mu \to e
\gamma$ process.

We find values of the same order for the partial amplitudes
$A_n^L,\ A_n^R$ and $A_c^R$ of
eq.~(\ref{ampl}) that contribute to the decay in eq.~(\ref{rate}). $A_c^L$ 
arises due to Yukawa interactions and
is roughly three orders of magnitude smaller than the other  
amplitudes. The relative sign of the amplitudes depends on the texture
and values of the input parameters. We observe
accidental cancellations of
partial amplitudes for values of $\sin\theta$ close to 1. Still, 
the effect is not very important, since the amplitudes are 
very small for these
values of  $\sin\theta$ and therefore hard to observe in Fig.2. 
The flavor mixing elements introduced by $\delta_{LL}^l$ 
and  $\delta_{RR}^l$ (\ref{Deltas})
in the scalar matrices, have a larger impact on the amplitudes than those
due to the trilinear terms $\delta_{LR}^l$. 

\section{{\large \bf $BR(l_j\to l_i \gamma)$ in the type I string model}}

The structure of the soft-terms in the type I string model is more
complicated than in the previous model. They depend, in addition
to  $m_{3/2}$ and $\theta$, on the values of the parameters
$\Theta_1$, $\Theta_2$ and $\Theta_3$. However, the flavor structure of the
slepton matrices is simpler, since the soft masses for the left--handed
sleptons of eq.~(\ref{scalarBL}) are universal at the GUT scale and 
the sneutrino mass
matrix of eq.~(\ref{sneutrino}) remains diagonal under a
rotation that diagonalizes
$Y_l$.  Therefore diagram (b) in Fig.1 does not contribute to the 
LFV processes calculated here.

The restrictions imposed by the experimental limits on the searches
for charginos and sleptons will constrain the free parameters of the model.
A  bound of $m_{\tilde {\chi}^+}=95 \ \rm{GeV}$ is found to be the most
severe on the initial conditions of eqs.~(\ref{m1}) and (\ref{m2}).
The restrictions
imposed by the bound of 90 GeV on the mass of the lightest charged
slepton will impose constraints on eqs.~(\ref{scalarBL}) and (\ref{scalarBR}).
These constraints depend on the values of the $\Theta_i$'s
as we will discuss later.

The predictions we obtain with this
model for  $BR(l_j\to l_i \gamma)$ allow us to simplify our 
presentation by setting
$\Theta\equiv\Theta_1=\Theta_2$. In the case of texture I, this
is justified by the fact that the experimental
bound on $BR(\mu\to e \gamma)$ tolerates a small deviation 
of the $\Theta_i$~'s from the common value of $1/\sqrt{3}$,  for 
which the soft masses become universal
(see Fig. 3). For the case of texture II, these predictions are 
more tolerant
to a variation of the $\Theta_i$~'s. However, when this texture is 
considered,  the
experimental
limit on $BR(\tau\to \mu \gamma)$ is more restrictive
(see Figs.~4, 5 and 6), since this bound is particularly
sensitive to the value $\Theta_3$. Therefore, we find that by setting also
 $\Theta_1=\Theta_2$ in the analysis of our results with texture II 
we can achieve a clearer presentation without any loss of generality.

Fig.~3 shows the constraint imposed by the current bound on the
$BR(\mu\to e \gamma)$ on the plane ($\sin\theta-\Theta$) for
constant values of $m_{3/2}=200 \ \rm{GeV}$ (left) and
$m_{3/2}=400 \ \rm{GeV}$ (right) when texture I is assumed.
Fig.4 displays  the equivalent for texture II. The light shaded areas
correspond to the space of parameters allowed by the bounds on
the masses of the SUSY particles. The region below the upper
dashed line corresponds to values of
 $m_{\tilde {\chi}^+}> 95 \ \rm{GeV}$,  while the sector above the lower
solid line corresponds to values of the lightest charged scalar
$m_{\tilde {l}}> 90 \ \rm{GeV}$.

The shape of the curve  $m_{\tilde {l}}= 90 \ \rm{GeV}$ in Figs.~3 and 4 is
determined by the initial conditions
given by eqs.~(\ref{scalarBL}) and  (\ref{scalarBR}). The lowest 
values for these masses corresponds 
to  $m_{R_1}=m_{R_2}$ when $1/\sqrt{3}<\Theta<1/\sqrt{2}$, while 
for  $\Theta<1/\sqrt{3}$, $m_{R_3}$ is
the lowest value. Therefore the largest component of the lowest
eigenvalue of the charged slepton mass is the $\tilde{e}_R$ or
the $\tilde{\tau}_R$ depending on the ranges of $\Theta$ above. Similar
considerations explain the different shape of
the curves for $m_{\tilde {l}}= 170 \ \rm{GeV}$ (with $m_{3/2}=200 \ \rm{GeV}$
and $m_{3/2}=400 \ \rm{GeV}$).

The darkest dotted areas in  Figs.~3 and 4 represent the sector
of parameters for which the lightest supersymmetric particle (LSP)
is a charged slepton. For $m_{3/2}=200 \ \rm{GeV}$ these areas are
below the bound of
$m_{\tilde {l}}= 90 \ \rm{GeV}$. However for  $m_{3/2}=400 \ \rm{GeV}$, the
cosmological requirement on the LSP to be a neutral particle (lightest
neutralino in our case) imposes a further restriction on the space of
parameters of the model.

Similarly to the results found for the WCHS model,
the assumption of  texture I for $Y_l$ allows a small deviation from the
universality of the scalar masses once we impose the experimental bound
on  $BR(\mu\to e \gamma)$ (light dotted sector inside of the grey
area  in Fig. 3). However
we found that the corresponding limit on  $\tau\to \mu \gamma$ does not
constraint the space of parameters shown in Fig. 3. The fact that the
branching ratios decrease with $m_{3/2}$ is reflected in a wider
light dotted area on the graphic corresponding to $m_{3/2}=400 \ \rm{GeV} $
in Fig.~3.

The flavor mixing elements in the scalar charged scalar mass matrix
are introduced by $\delta_{RR}^l$  and  $\delta_{LR}^l$ in
eq.~(\ref{Deltas}). As stated before, only diagram (a) from
Fig.~1 contributes to the rare lepton decays under consideration.
The non--universality of the right sleptons of
eq.~(\ref{scalarBR}) enhances the partial amplitude $A^L_{n}$ in
eq.~(\ref{ampl}) over the  $A^R_{n}$ which is due to the mixing
$\delta_{LR}^l$ originated from the flavor dependent structure of
the A-terms of eq.~(\ref{AtermB1}). The values we find for
$A^R_{n}$ are of the order of  10\% of the values of $A^L_{n}$. We 
stress that a particular choice of
the $\Theta_i$'s such that $|\Theta_i|=1/\sqrt{3}, \ i=1,2,3$
with $\Theta_2$ or $\Theta_3$ negative, will maintain the
universality of the scalar masses while producing a maximal
non--universality in the A--terms (see eq.~(\ref{AtermB2})). In
this particular case the mixing introduced by  $\delta_{LR}^l$ can
induce branching ratios of the order of the ones here presented.

When texture~II is assumed for $Y_l$, the phenomenological
constraints imposed by SUSY particles on the parameter space of
the model do not differ significantly from the case of texture~I.
Lines corresponding to constant masses for sleptons and charginos
in Fig.4 would be located in the same places as in Fig.~3. The
present limits on  $BR(\mu\to e \gamma)$ do not impose any
restriction on the parameter space shown in Fig.~4. However, in
the case of texture~II, the limits on the decay  $\tau\to \mu
\gamma$ restrict the space of parameters to the dotted region inside of
the gray area on the graphics of Fig.~4. As one can immediately see, this
constraint is significant for the case of $m_{3/2}=200 \
\rm{GeV}$ and decreasingly restrictive as $m_{3/2}$ increases to $400
\ \rm{GeV}$. From these figures one  can induce the effect
of improving the present experimental limits on the two processes
considered.

Fig. 5 shows the behavior of the branching ratios with $\sin\theta$ for
fixed values of $\Theta$, for the case of  texture II 
and $m_{3/2}=200 \ \rm{GeV}$. These
ratios decrease as $\sin\theta$ increases since higher values of  this
parameter are related to larger values of the masses 
of the supersymmetric
particles present on diagram (a) of Figure 1. As the values of
$\Theta$ get closer to $1/\sqrt{3}$ the prediction for the ratios
is smaller and will eventually vanish for the limit of universal soft
terms corresponding to $\Theta=1/\sqrt{3}$. We observe that the
allowed range of $\sin\theta$ is larger for values of $\Theta>1/\sqrt{3}$
(solid lines), than for values of  $\Theta<1/\sqrt{3}$, which is
explained by the asymmetry of the shaded  regions on Figs.~3 and 4.

Fig. 6, in a complementary way to Fig. 5,
shows the behavior of the branching ratios with $\Theta$ for
fixed values of $\sin\theta$ also for the case of  texture II 
and $m_{3/2}=200 \ \rm{GeV}$. As one can see, the universality 
of the soft terms obtained for $\Theta=1/\sqrt{3}$ leads to 
vanishing ratios. As
the values from  $\Theta$ differ from this value the ratios increase.
As we mentioned before, higher values of $\sin\theta$ correspond to lower
values of the ratios. The ranges on $\Theta$ for different values of
$\sin\theta$ can be understood from the allowed space of parameters 
represented on Figs.~3 and 4.

\section{{\large \bf Conclusions}}
We have studied the predictions for the LFV decays $\mu \to e
\gamma$ and  $\tau \to \mu \gamma$ arising from 
non universal soft terms as they appear when the MSSM
is derived from a general string theory. We studied the dependence 
of these predictions on a general, non--diagonal,  texture of the lepton 
Yukawa couplings.

The results found show the relevance of the considered processes in
constraining the undetermined parameters of the models, and 
therefore their predictions for the SUSY particles. On the other hand,
these processes can provide some information on the charged 
lepton Yukawa couplings, which can be very important when this 
models are extended to explain neutrino physics. 


We found the  non--universality of the soft masses to be more
relevant for LFV than those of the $A$-term are. However, the latter
ones are of phenomenological interest for other processes such as 
CP violation effects. 
Finally, we would like to emphasize the importance of the improvements on the
current experimental limits on LFV processes to understand the
nature of the flavor problems on the SUSY extensions of the 
SM. 

\section*{\large{\bf Acknowledgements}}
Our research has been supported by the E.U. T.M.R. Network
contract ERBFMRX-CT96-0090 (M.E.G.), PPARC (S. K.) and F.C.T. 
PRAXIS XXI/BD/9416/96 (D.F.C.). 
We would like to thank J.C. Rom\~ao for his help in the 
programing stages of our work  
and for many useful discussions. We also thank  D. Bailin 
and  A. M. Teixeira for carefully reading the manuscript.

\section*{\large{\bf Note added in proof}}
After the first version of this work was completed the Muon $(g-2)$ 
Collaboration had published a new measurement of the anomalous 
magnetic moment of the muon 
(H. N. Brown {\it et al.}, hep-ex/0102017). The result presented differs 
from the SM prediction by 2.6$\sigma$. The MSSM  
contribution to	this process, excludes one of the signs of the 
$\mu$--parameter ($\mu<0$ in our convention) for the space of 
parameters described in this work. The results we presented 
correspond to $\mu>0$. However, we should indicate that the a 
change of sign of the 
$\mu$--parameter parameter do not change significantly our
predictions the branching ratios.


\newpage

\section*{Figures}

\begin{figure}[h]
\centerline{
\epsfig{figure=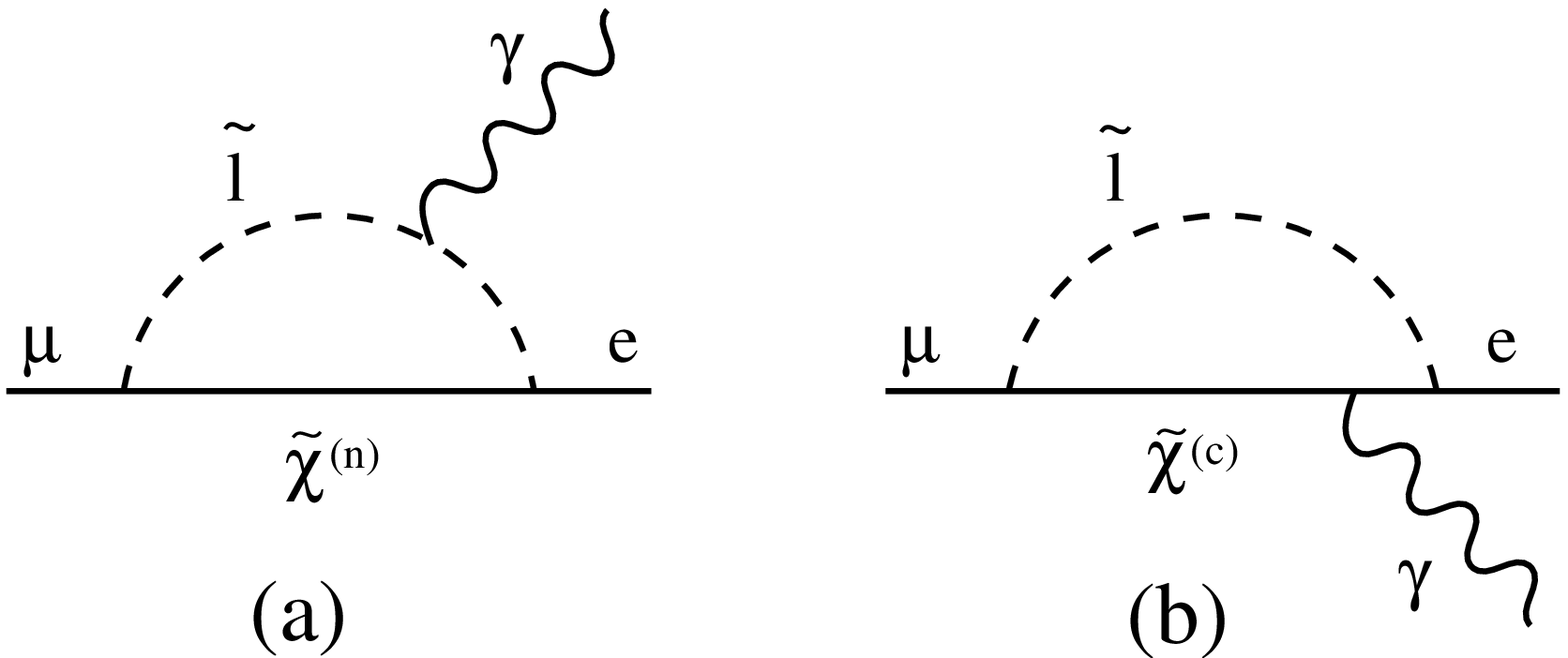,height=5cm,width=10cm,angle=0}\\
}
\vskip .5cm
\caption{The generic Feynman diagrams for $\mu \to e \gamma$ decay.
$\tilde{l}$ stands for charged slepton (a) or sneutrino (b), while
${\tilde\chi}^{(n)}$ and ${\tilde\chi}^{(c)}$ represent
neutralinos and charginos
respectively.}
\end{figure}

\begin{figure}[b]
\hspace*{-0.5in}
\begin{minipage}[b]{9in}
\epsfig{figure=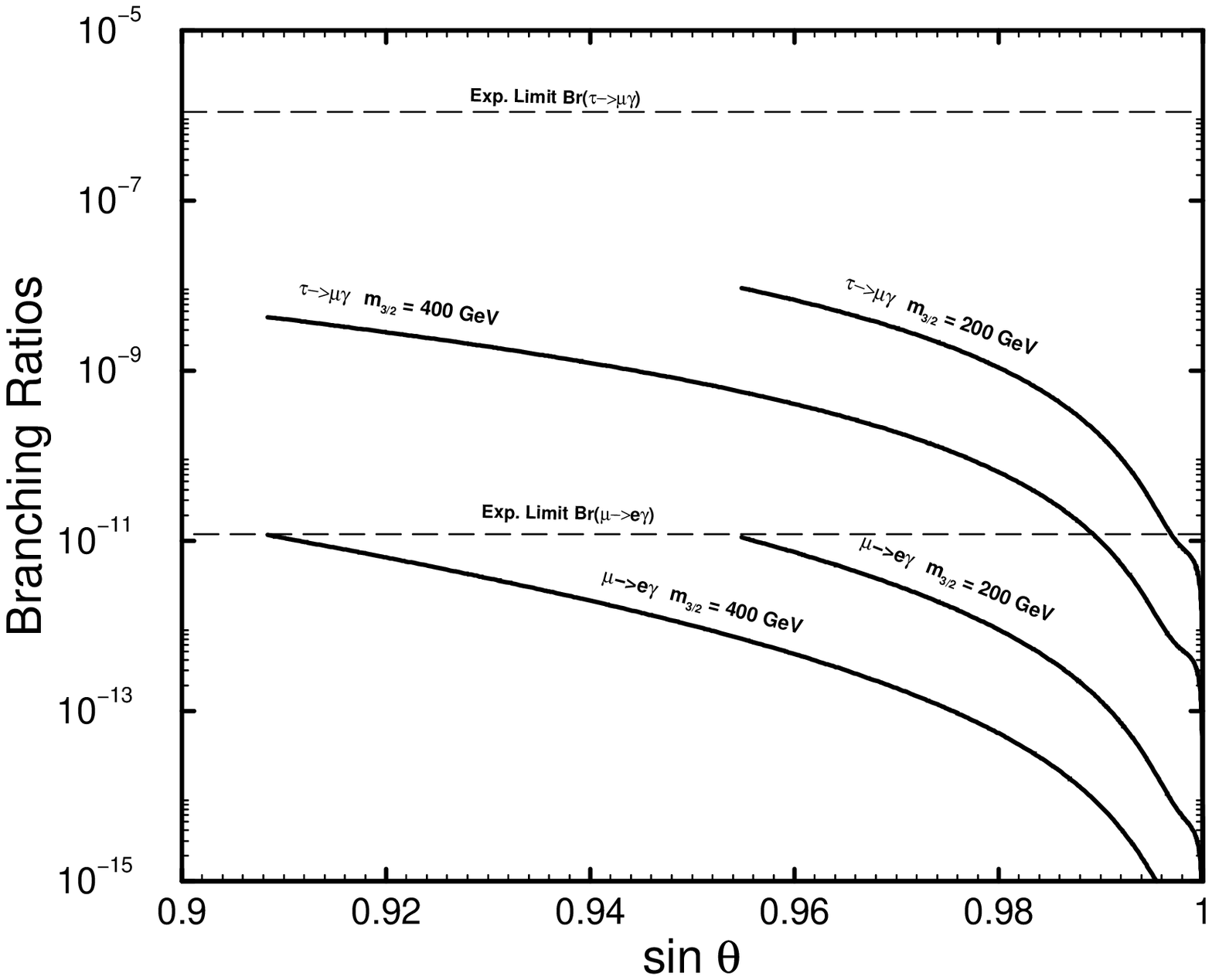,height=3.2in,width=3.2in,angle=0}
\epsfig{figure=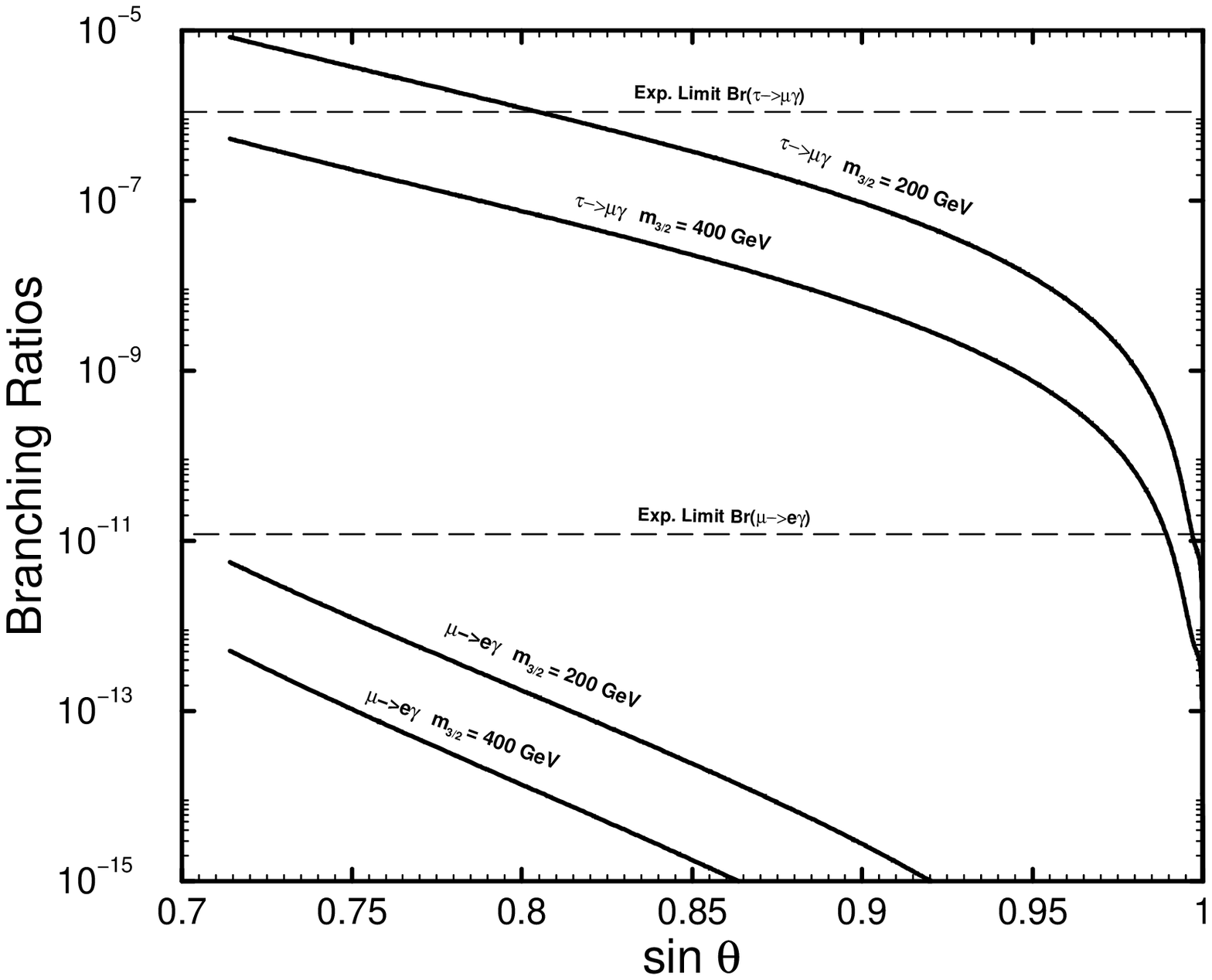,height=3.2in,width=3.2in,angle=0}
\end{minipage}
\medskip
\caption{Branching ratios vs. $\sin\theta$ for the WCHS model with
texture I for $Y_l$ (left) and texture II (right) and $\tan\beta=10$. 
The values for $m_{3/2}$ are kept constant as shown on the curves.
\label{fig.2}}
\end{figure}
\begin{figure}
\hspace*{-0.5in}
\begin{minipage}[b]{9in}
\epsfig{figure=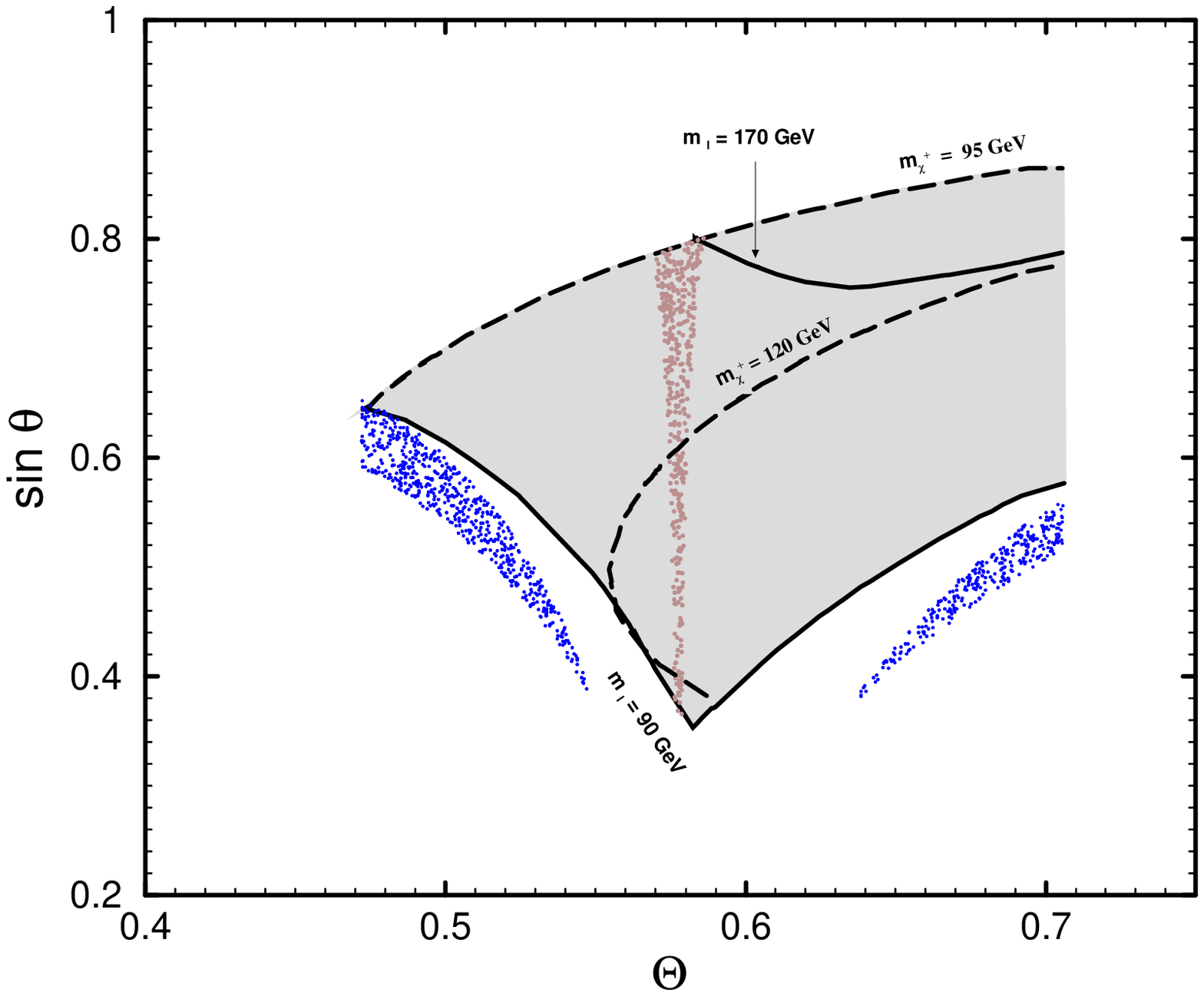,height=3.2in,width=3.2in,angle=0}
\epsfig{figure=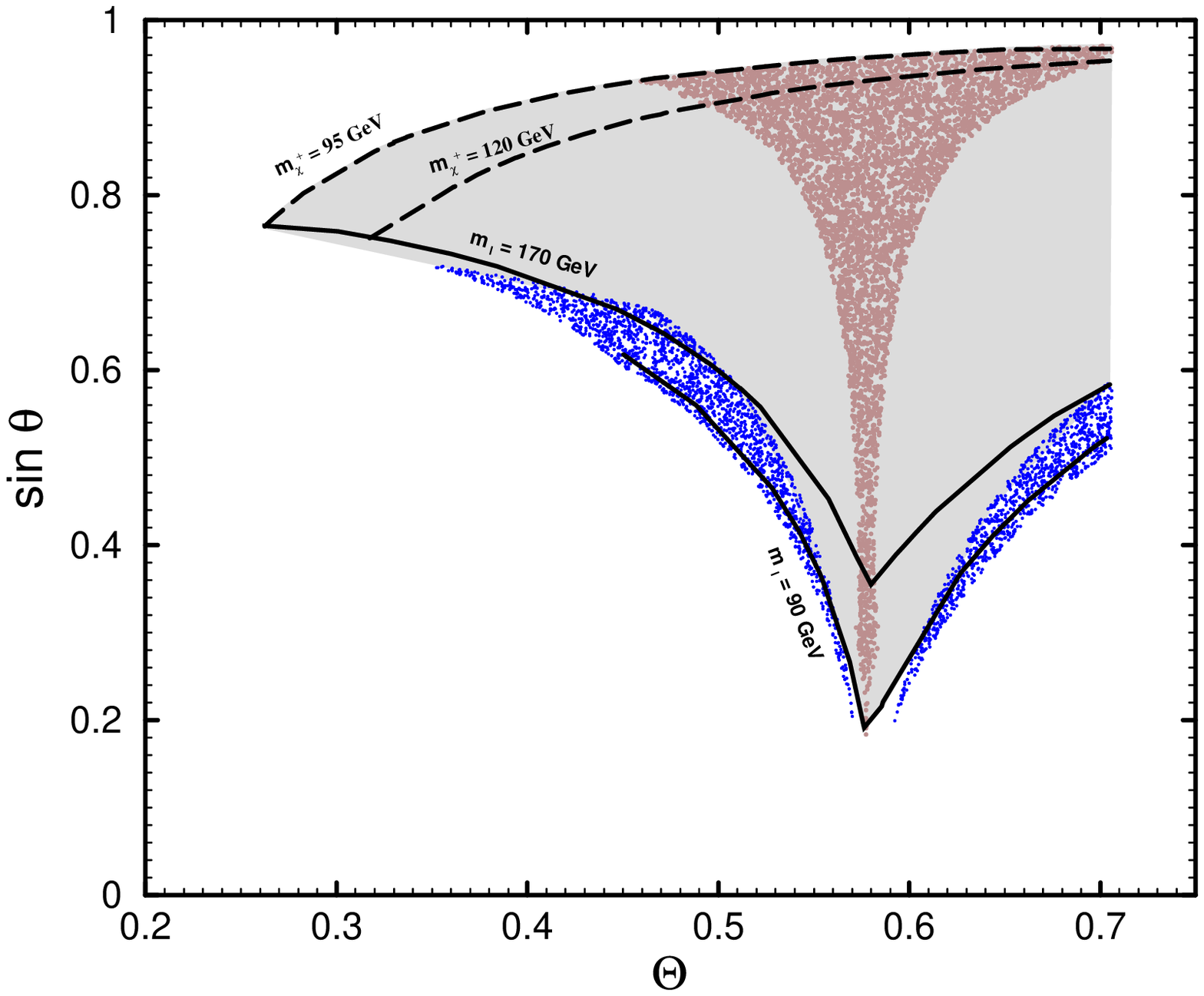,height=3.2in,width=3.2in,angle=0}
\end{minipage}
\medskip
\caption{Areas with $BR(\mu\to e \gamma)< 1.2 \times 10^{-11}$
(dotted areas inside the gray contour) in the plane
$\sin\theta-\Theta$ for constant values of
$m_{3/2}=200\ \rm{GeV}$ (left) and
$m_{3/2}=400\ \rm{GeV}$ (right) and $\tan\beta=10$.
The model used corresponds to type I string, with texture I for $Y_l$.
Values of the masses of SUSY particles which bound the parameter
space of the model are as shown in the graphs. The dark dotted areas
correspond to space of parameters such that the LSP is a slepton.
\label{fig.3}}
\end{figure}

\begin{figure}
\hspace*{-0.5in}
\begin{minipage}[b]{9in}
\epsfig{figure=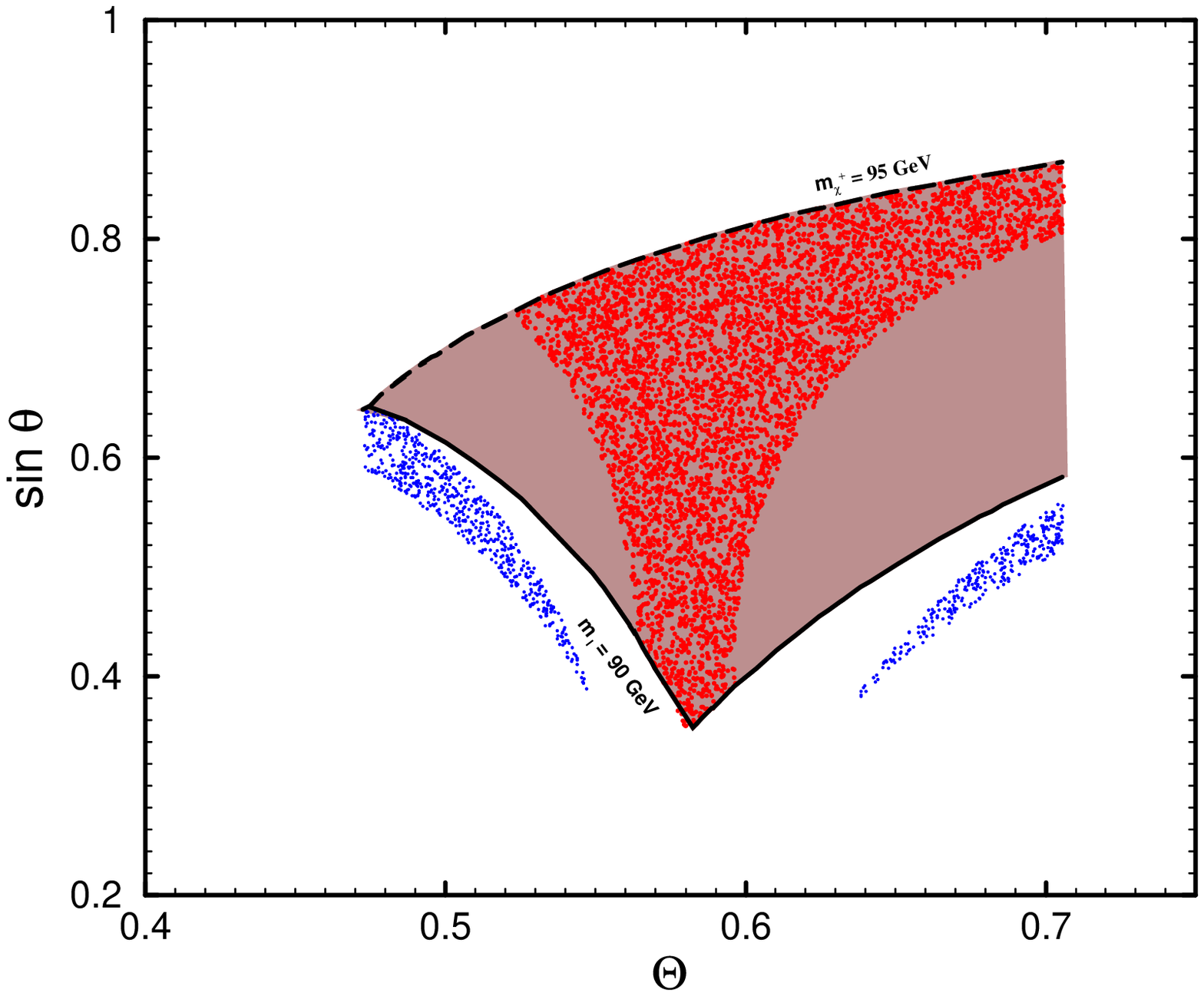,height=3.2in,width=3.2in,angle=0}
\epsfig{figure=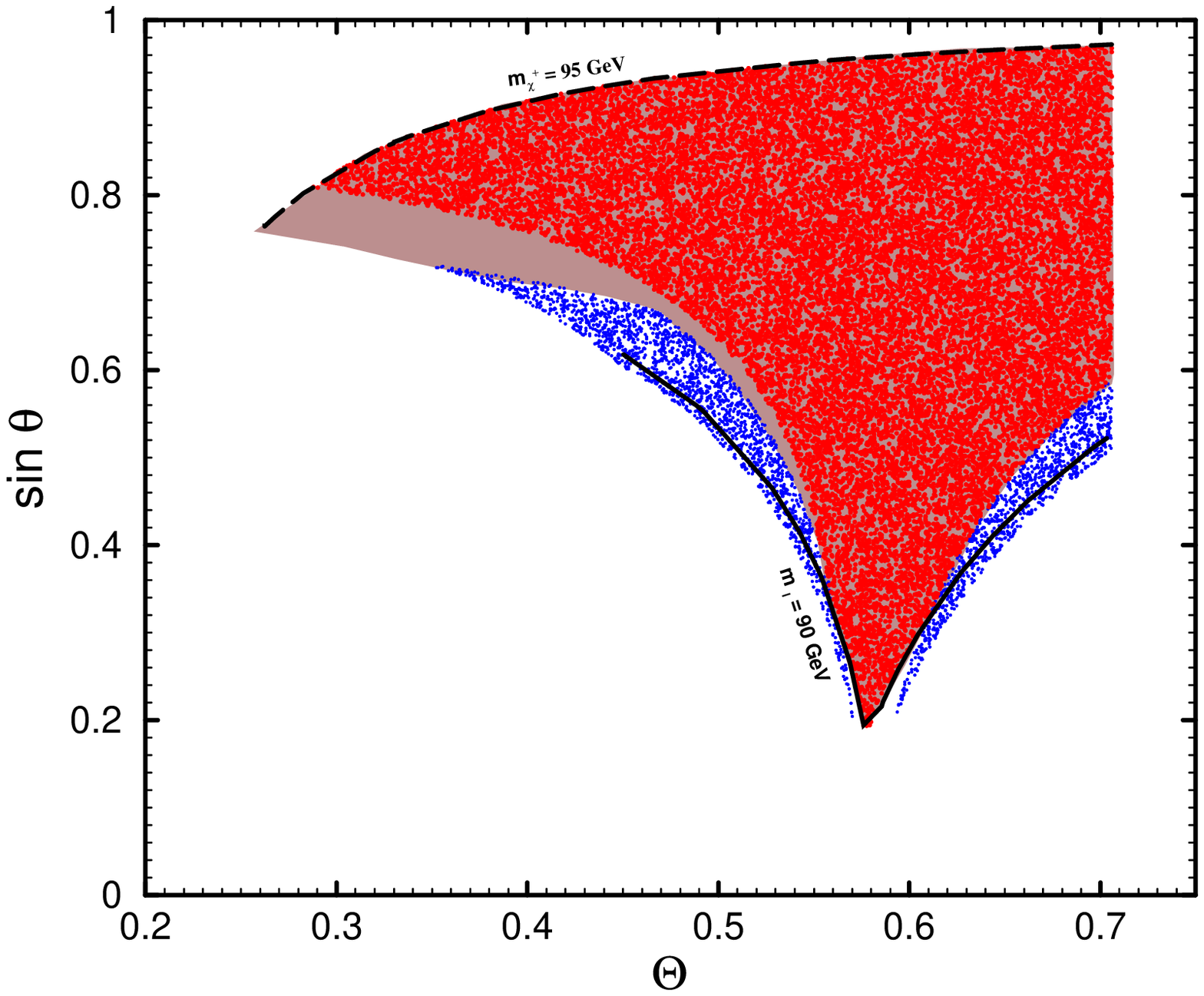,height=3.2in,width=3.2in,angle=0}
\end{minipage}
\medskip
\caption{Areas with $BR(\tau\to \mu \gamma)< 1.1 \times 10^{-6}$
(dotted areas inside the gray contour) in the plane
$\sin\theta-\Theta$ for constant values of
$m_{3/2}=200\ \rm{GeV}$ (left) and
$m_{3/2}=400\ \rm{GeV}$ (right) and  $\tan\beta=10$.
The model used corresponds to type I string, with texture II for $Y_l$.
Values of the masses of SUSY particles which bound the parameter
space of the model are as shown in the graphs. The dark dotted areas
correspond to space of parameters such that the LSP is a slepton.
\label{fig.4}}
\end{figure}

\begin{figure}
\hspace*{-0.5in}
\begin{minipage}[b]{9in}
\epsfig{figure=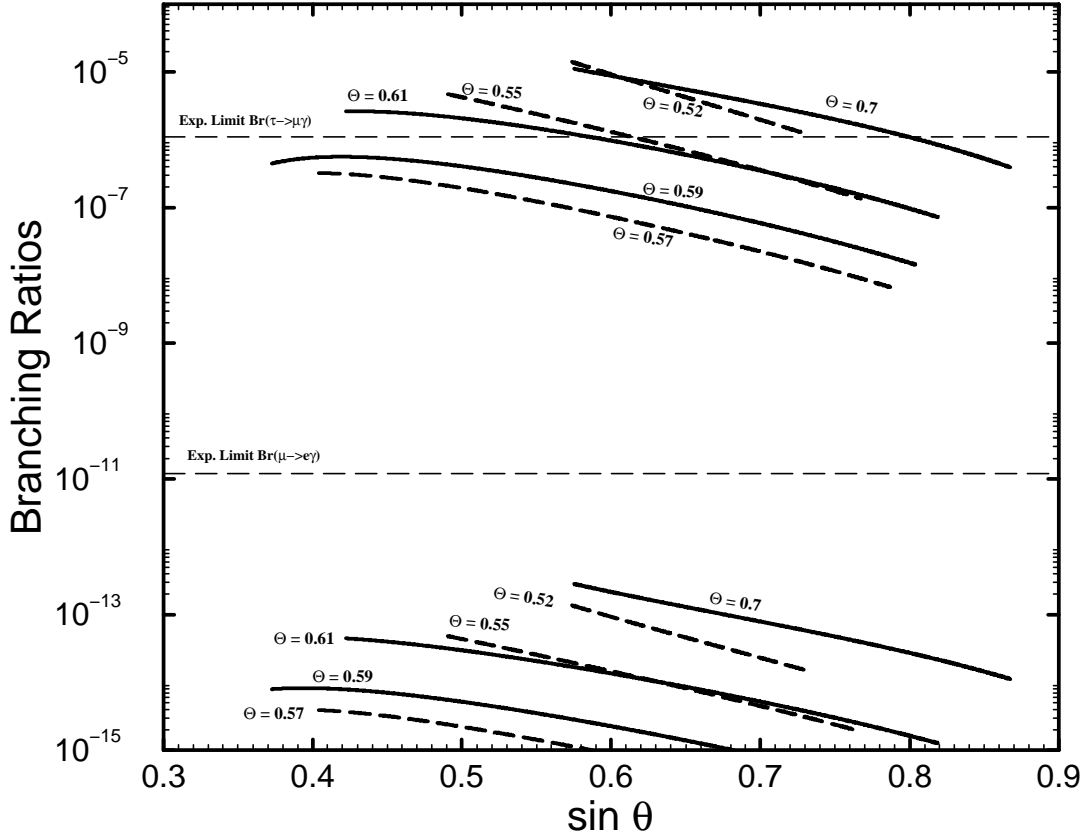,height=4.5in,angle=0}
\end{minipage}
\medskip
\caption{Branching ratios vs. $\sin\theta$ for the type I string
model with texture II for $Y_l$, 
$m_{3/2}=200$ GeV and $\tan\beta=10$. The values 
for $\Theta$ are kept constant as
shown on the curves, solid (dashed) lines correspond to values for
$\Theta>1/\sqrt{3}$($\Theta<1/\sqrt{3}$).
\label{fig.5}}
\end{figure}

\begin{figure}
\hspace*{-0.5in}
\begin{minipage}[b]{9in}
\epsfig{figure=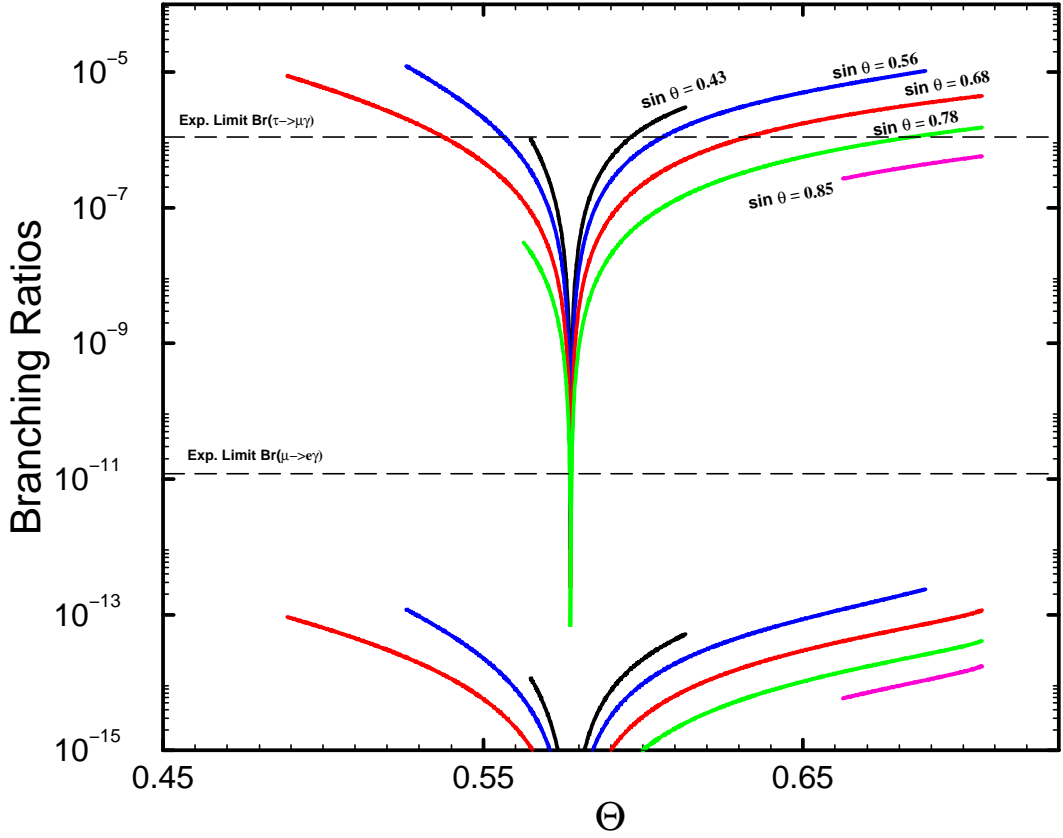,height=4.5in,angle=0}
\end{minipage}
\medskip
\caption{Branching ratios vs. $\Theta$ for the type I string
model with texture II for $Y_l$, 
$m_{3/2}=200$ GeV and  $\tan\beta=10$. The values 
for $\sin\theta$ are kept constant as
shown on the curves.
\label{fig.6}}
\end{figure}

\end{document}